# A Longitudinal Analysis about the Effect of Air Pollution on Astigmatism for Children and Young Adults

Lin An[1], Qiuyue Hu[1], Jieying Guan[2,3], Yingting Zhu[2,3], Chenyao Jiang[1], Xiaoyun Zhong[1], Shuyue Ma[1], Dongmei Yu[4], Canyang Zhang[1], Yehong Zhuo[2,3*], Peiwu Qin[1*]

[1]Tsinghua-Berkeley Shenzhen Institute, Institute of Biopharmaceutical and Health Engineering, Tsinghua University, Tsinghua Park, Xili University Town, Nanshan District, Shenzhen, Guangdong, China

[2]State Key Laboratory of Ophthalmology, Zhongshan Ophthalmic Center, Sun Yat-sen University, Guangdong Provincial Key Laboratory of Ophthalmology and Visual Science, Guangzhou, China

[3]Guangdong Provincial Clinical Research Center for Ocular Diseases, Guangzhou, China

[4]School of Mechanical, Electrical & Information Engineering, Shandong University, Weihai, Shandong, China

First Authors: Lin An, Email: al20@mails.tsinghua.edu.cn, Tel.: (+86)13528416992; Qiuyue Hu, Email: qy-hu20@mails.tsinghua.edu.cn, Tel.: (+86) 13719240134. These authors have contributed equally to this work.

Corresponding Author: Peiwu Qin, Email: pwqin@sz.tsinghua.edu.cn, Tel.: (+86)19910528642, Fax: 755-36882133, No. 2279, Lishui Road, Nanshan District, Shenzhen 518055, China. Yehong Zhuo,

Email: zhuoyh@mail.sysu.edu.cn, Tel:(+86)13352828998, Fax: 020-66618962, No. 7, Jinsui Road, Tianhe District, Guangzhou 510060, China; These authors contributed equally to this work.

Word count (not including title page, acknowledgment, references, and figure legends): 3404 words

Funding: This research was supported by the National Key Research and Development Project of China (2020YFA0112701); the National Natural Science Foundation of China (82171057); Science and Technology Program of Guangzhou, China (202102010216); National Natural Science Foundation of China (31970752); Science, Technology, Innovation Commission of Shenzhen Municipality (JCYJ20190809180003689, JSGG20200225150707332, JSGG20191129110812708, ZDSYS20200820165400003, WDZC20200820173710001); Shenzhen Bay Laboratory Open Funding (SZBL2020090501004).

# Abstract

**Purpose**: This study aimed to investigate the correlation between air pollution and astigmatism, considering the detrimental effects of air pollution on respiratory, cardiovascular, and eye health.

**Methods**: A longitudinal study was conducted with 127,709 individuals aged 4-27 years from 9 cities in Guangdong Province, China, spanning from 2019 to 2021. Astigmatism was measured using cylinder values. Multiple measurements were taken at intervals of at least 1 year. Various exposure windows were used to assess the lagged impacts of air pollution on astigmatism. A panel data model with random effects was constructed to analyze the relationship between pollutant exposure and astigmatism.

**Results**: The study revealed significant associations between astigmatism and exposure to carbon monoxide (CO), nitrogen dioxide ($NO_2$), and particulate matter ($PM_{2.5}$) over time. A 10 μg/m³ increase in a 3-year exposure window of $NO_2$ and $PM_{2.5}$ was associated with a decrease in cylinder value of -0.045 diopters and -0.017 diopters, respectively. A 0.1 mg/m³ increase in CO concentration within a 2-year exposure window correlated with a decrease in cylinder value of -0.009 diopters. No significant relationships were found between $PM_{10}$ exposure and astigmatism.

**Conclusion**: This study concluded that greater exposure to $NO_2$ and $PM_{2.5}$ over longer periods aggravates astigmatism. The negative effect of CO on astigmatism peaks in the exposure window of 2 years prior to examination and diminishes afterward. No significant association was found between $PM_{10}$ exposure and astigmatism, suggesting that gaseous and smaller particulate pollutants have easier access to human eyes, causing heterogeneous morphological changes to the eyeball.

**Introduction**

Astigmatism is a common type of refractive error caused by an imperfect curvature in the eye's cornea or lens, impacting different aspects of visual development. First, astigmatism may cause a blurry image on the retina, disrupting emmetropization [1, 2]. Second, uncorrected astigmatism can lead to meridional amblyopia in those at the stage of vision development [3-5]. Finally, this type of refractive error will seriously impair the school children's learning ability [6] and hinder their physical and mental development [7]. Advanced imaging technologies and the invention of machine learning algorithms were applied to improve accuracy of detection[8-13]. Investigating the risk factors for astigmatism is crucial for refractive intervention in children and young adults. Previous studies have compared the prevalence of refractive errors in urban and rural areas, revealing a higher myopia rate among urban students [14-16]. Given the perceived difference in air pollution exposure between urban and rural contexts, it is important to examine the association between air pollution and astigmatism [17].

As one of our era's most significant health threats, air pollution is composed of particulate matter (PM), carbon monoxide (CO), ozone ($O_3$), nitrogen dioxide ($NO_2$), and sulfur dioxide ($SO_2$). Numerous studies have linked poor air quality to higher rates of morbidity and mortality across various diseases [18]. According to a 2013 meta-analysis, a 10 μg/m³ rise in annual particulate matter less than 2.5 μm in aerodynamic diameter ($PM_{2.5}$) concentration resulted in an 11% increase in cardiovascular mortality [19]. Moreover, higher ambient air pollution levels appeared to elevate the chance of being diagnosed with vascular dementia and Alzheimer's disease [20]. Air pollution's detrimental effects extend to the eyes, causing various ophthalmological diseases. Chang et al. [21] mentioned that the possibility of outpatient visits for nonspecific conjunctivitis was correlated with the contamination $NO_2$, $SO_2$, $O_3$, and particulate matter less than 10 μm in aerodynamic diameter ($PM_{10}$). Adar et al. [22] established the association between higher air pollution concentrations and narrower retinal arteriolar diameters, which suggested that higher exposures to air pollutants were linked to cardiovascular morbidity and mortality. Chua et al. [23] ascertained the adverse influence of $PM_{2.5}$ upon glaucoma and macular ganglion cell-inner plexiform layer (GCIPL). However, the relationship between air pollution and astigmatism remains elusive.

The cross-sectional design is often adopted to evaluate the health impacts of air pollutants [22-25]. Nevertheless, such a snapshot at a single timepoint may not capture the impacts over time because the damage caused by air pollution to the human body is a long-term, lagged, and chronic process [23]. Longitudinal research designs collect data over time, allowing for a better understanding of the effects of historical exposure to air pollution [24]. Few studies exploit this method to figure out if ocular diseases or vision disorders are linked to air pollution.

In this study, we used a large dataset from 9 cities in Guangdong Province, China, where each participant had at least 2 astigmatism measurements taken at ≥1-year intervals to analyze the associations with air pollution. To the best of our knowledge, this is the first study to investigate the relationships between astigmatism and air pollution. We implemented a panel data model, a configuration of longitudinal design to take into account the lagged effects of air pollution on astigmatism progression. This study provides a novel, potentially modifiable risk factor for refractive intervention and would arouse more attention to air pollution prevention.

**Methods**

**Ethics Statement**

This school-based longitudinal study was approved by the Institutional Review Board/Ethics Committee of Sun Yat-sen University, Guangzhou, China (No. 2021KYPJ185), and registered with the Chinese Clinical Trial Registry (ChiCTR) (identifier: ChiCTR2200057391). Consent was waived because all datasets were de-identified before transfer for data analysis.

**Study Population**

Guangdong Province, located in southeastern coastal China, comprises 21 cities with a total area of approximately 179,700 km² and a population of 115.21 million as of 2021. This ocular-related data was collected from 9 cities (Guangzhou, Heyuan, Huizhou, Qingyuan, Shantou, Shanwei, Shaoguan, Yunfu, and Zhuhai) in Guangdong Province between 2019 and 2021. The dataset contained ocular information from 127,709 children and young adults aged 4-27 years, and each individual had ≥2 observations at ≥1-year intervals.

**Data collection and exclusion criteria**

The dataset for this study was collected using an online platform provided by Guangdong Eyevision Medical Technology Co., Ltd., a national high-tech enterprise specializing in medical technology services and health data standardization. The corporation has been conducting vision screening for school-aged children and young adults in Guangdong Province since 2018, in line with national policies on adolescent myopia prevention and control. The platform allows users to upload, store, and manage vision screening data, including information such as students' ID, name, gender, age, grade, school type, visual acuity, and cylinder power.

Prior to data collection, staff in the participating schools received training on screening preparation, including setting examination schedules, selecting suitable sites, creating vision health records, and inputting data into the platform. Detailed information about each student, such as identity card number (ID), age, gender, educational stage, school type, and city of residence, was collected through face-to-face interviews with the head teacher of each class. Optometrists who were experienced and certified performed the ocular examinations according to standard protocols. The cylinder power, representing the amount of lens power for astigmatism, was measured in each eye using a Topcon KR-1 device. The average of three reliable measurements was taken as the final result. Information on wearing glasses and the type of glasses worn was also recorded during the examinations.

During the recruitment period, a total of 868,222 observations from 416,048 participants were collected from 1,659 schools. Some observations were excluded due to missing data on wearing glasses, cylinder power, or city of residence, as well as participants who changed their city of residence or reported pre-existing ocular conditions. The final dataset included 269,963 observations from 127,709 participants for analysis.

**Air Pollution Measurement**

We obtained city-wide average daily concentrations of CO, $NO_2$, $PM_{2.5}$, and $PM_{10}$, from 2016 to 2021 from China National Environmental Monitoring Centre, which gathered the historical air quality of Chinese cities. Following studies that used average pollutant concentrations of exposure window as the measure of cumulative exposure [26, 27], we defined four exposure windows of different lengths, including the year of examination and 1, 2, and 3 years prior. By matching participants' residence city, examination date, and selected window length, we calculated the estimated long-term effects of pollutants on each participant using the average pollutant concentrations during the corresponding period.

**Statistical Analysis**

We aim to identify a suitable model to examine the lagged effect of air pollutants on astigmatism across different exposure time lengths. The linear mixed model (15) and the panel data model are considered appropriate for longitudinal studies, with the former sharing assumptions with the standard linear regression model [28]. However, the small p-values acquired from White-Test (p=0.007) and Breusch-Pagan-Test (p=0.0012) indicate heteroskedasticity and a result of 1.16 from the Durbin-Watson-Test signifies that there is a certain degree of positive autocorrelation, which all indicate that the linear mixed model is not suitable for our data. In contrast, the panel data model incorporates both cross-sectional and time series information [27]. It captures both individual differences at a specific time point and individual changes over time. A typical panel data model for individual i = 1, …, n with observations collected at multiple time points t = 1, …, T takes the form that

$$y_{it} = \beta^T x_{it} + u_i + e_{it}$$

where $y_{it}$ is the dependent variable, $\beta$ is an M-dimensional column vector of parameters, $x_{it}$ is an M-dimensional column vector of predictors, $u_i$ is the individual-specific effects and $e_{it}$ is an idiosyncratic error term. The panel data model includes three variants: the pooled Ordinary Least Squares (OLS), the random effects (RE), and the fixed effects (FE) models. The pooled OLS model is a basic OLS model commonly used as a reference for comparing other models' performance. The RE model assumes that the individual-specific effect $u_i$ is unrelated to the predictors $x_{it}$, whereas the FE model claims the opposite [29]. These variants can be further separated into two broad categories: homogeneous (or pooled) panel data models and heterogeneous models [30]. The latter involving fixed effects and random effects models allow for heterogeneity across groups and introduce the individual-specific effects.

To determine the model that best integrates the data characteristics and speculates parameters accurately, we need to test the significance of the FE and RE using the F-test and Breusch-Pagan Lagrange Multiplier (LM) test, respectively. The null hypothesis of the F-test is that all individual-specific effects are jointly zero. The test statistic for the F-test is calculated as follows:

$$F - statistic = \frac{(SSE_r - SSE_u)/(k_2 - k_1)}{SSE_u/(N - k_2)}$$

where $SSE_r$ is the error sum of squares from the restricted model (pooled OLS model) and $SSE_u$ is the error sum of squares from the unrestricted fixed effects model. $k_1$ is the degrees of freedom for the restricted model and $k_2$ is equal to the degrees of freedom for the unrestricted model. $N = n \times T$ is the number of data samples. The F-statistic obeys the F-distribution with $(k_2 - k_1, N - k_2)$ degrees of freedom. The F-statistic greater than the critical value at alpha $= .05$ means that the FE turns out significant.

To choose between the RE model and pooled OLS model, Breusch-Pagan Lagrange Multiplier (LM) test was used to test the significance of the RE model, whose null hypothesis is that the variance of the RE is zero. The LM test statistic which is Chi-squared (1) distributed is as follows:

$$LM = \frac{nT}{s(T-1)} \left[ \frac{T^2 \sum_{i=1}^{n} \overline{\epsilon}_i}{\sum_{i=1}^{n} \sum_{t=1}^{T} \epsilon_{it}} - 1 \right]^2$$

where n is the number of individuals, T is the number of time points, $\overline{\epsilon}_i$ represents the mean residual error from the pooled OLS model of individual i over all time points, and $\epsilon_{it}$ is the residual error from pooled OLS model of individual i at time point t. The test statistic of the LM test larger than the Chi-squared (1) critical value signifies the goodness-of-fit of the RE model is better than that of the pooled OLS model.

Provided that there are both FE and RE, the Hausman test, whose null hypothesis is that the (fixed or random) effects are independent of the predictors, would compare the two models and determine the better specification. Under the null hypothesis, the fixed effects model is consistent but inefficient, whereas the random effects model is both consistent and efficient. Failure to reject the null hypothesis favors the random effects specification [31].

Therefore, we constructed the models in combination with the above three tests to determine each air pollutant's lagged, historic influence on astigmatism. The F-test and LM test were used to identify the presence of FE or RE in the dataset. If both effects were significant, the Hausman test was used to determine the preferred model. The results of these tests were summarized in Supplementary Table S1.

The F-statistic values for all air pollutants' exposure windows were greater than the critical value which was 1.00 at alpha = 0.05, indicating the presence of fixed effects in all cases. Additionally, the LM statistic values exceeded the chi-squared critical value of 3.84 at alpha = 0.05, indicating significant random effects across all exposure windows. Finally, the Hausman test yielded p-values greater than or equal to 0.05 for all cases, suggesting that the random effects model was a better fit for our dataset.

According to the participant's ocular examination date, we chose the year of the examination and 1, 2, or 3 years before the examination date as the exposure windows (d = 0, 1, 2, 3). We calculated the average air pollution concentrations over the exposure windows as the lagged effects. Our model (Equation 1) for participant i = 1, ..., 127,709 who was examined in year t = 2019, 2020 or 2021 was as follows.

$$Y_{it} = \alpha + \beta_{0d}X_{itd} + \beta_{1d}^{T}Cov_{it} + u_{id} + e_{itd} \quad (1)$$

where $Y_{it}$ is participant i's cylinder power at ocular examination time t, α is the intercept, $\beta_{0d}$ is the main parameter of interest that estimates the association between air pollution concentration of the exposure window d and astigmatism, $\beta_{1d}$ is a column vector consisting of the coefficients of the covariates, $X_{itd}$ refers to the average air pollution concentration of exposure window preceding the examination date, $Cov_{it}$ is a column vector containing all time-varying covariates (school type, educational stage, school property and type of wearing glasses), $u_{id}$ is an individual-specific effect and $e_{itd}$ is an idiosyncratic error term.

We conducted two sensitivity analyses to check the robustness of our model. First, we examined the effect modification by gender through dividing the participants into two subgroups. Second, we investigated whether the impacts of air pollution varied with age in four groups separated by the quartile of the age, which were 4-10 years, 10-13 years, 13-15 years, and 15-27 years. For each air pollutant, we chose the exposure window which had the strongest association with cylinder as the baseline and re-ran our model adjusted for all covariates to evaluate the quantitatively difference in association as well as the changes in direction or in the statistical significance among diverse subgroups. We implemented our model using Python version 3.8.8 and linearmodels package version 4.24.

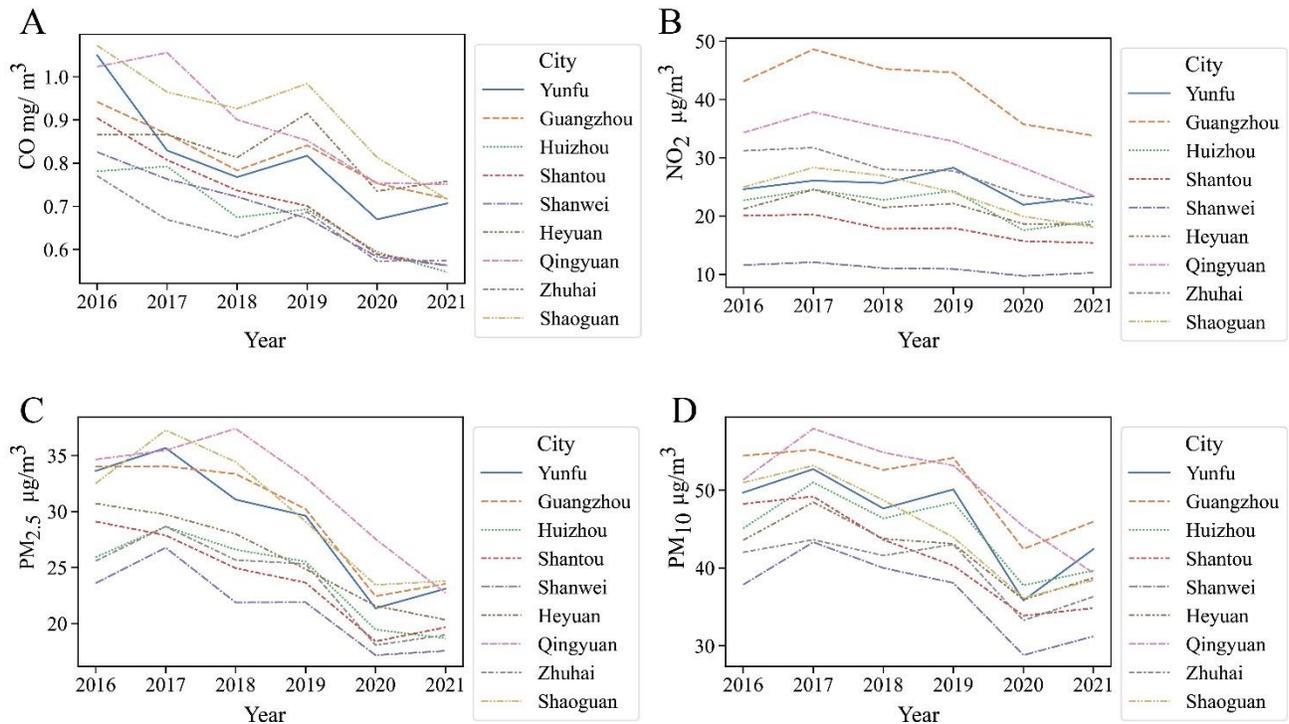

**Figure 1.** The annual average concentrations of CO, $NO_2$, $PM_{2.5}$ and $PM_{10}$ for 9 cities in Guangdong Province, China from 2016 to 2019. Note: CO, carbon monoxide; $NO_2$, nitrogen dioxide; $PM_{2.5}$, particulate matter less than 2.5 μm in aerodynamic diameter; $PM_{10}$, particulate matter less than 10 μm in aerodynamic diameter.

**Results**

Our study consisted of 127,709 individuals, and since each had 2 or 3 measurements at different years, there were 269,963 measurements in total. Table 1 shows the demographic characteristics of our study population, with an average age of 13 years old, approximately, and 52% being female. Most of the population did not wear glasses (69%), and about 30% of them wore frame glasses; 38% were in primary school, followed by 33% in junior high school, 23% in senior high school, 5% in university and 0.15% in kindergarten. The majority came from public schools (92%). 97% of the participants studied in general school, while the proportion in special school was the least, with only 14 measurements. The mean (SD) of the cylinder was -0.81 diopters (0.88 diopters).

Considering the earliest measurements all started in 2019, and we were interested in the lagged impact of air pollutants, we investigated the annual average concentrations of CO, $NO_2$, $PM_{2.5}$, and $PM_{10}$ for nine cities from 2016 to 2021, which were shown in Figure 1. Qingyuan had higher concentrations of all four pollutants than other cities, and Guangzhou displayed the highest level of $NO_2$. Figure 2 shows the maximum, minimum, and average levels of four air pollutants over the years.

Table 2 shows the associations between different exposure windows of air pollutants and the cylinder. Among the investigated air pollutants, CO, $NO_2$, and $PM_{2.5}$ were found to be related to the cylinder. Statistical significance was observed in all four windows (current, P1, P2, and P3) of $NO_2$. In comparison, it existed in three windows of CO (current, P1, and P2) and only one of $PM_{2.5}$ (P3). For CO, the average concentration of 2 years before the ocular examination had the most significant impact, with a -0.009-diopters decrease (95% CI, -0.014, -0.004) for a 0.1 mg/m³ concentration increase. The exposure window of 3 years before the examination for $NO_2$ demonstrated the strongest association that 10 μg/m³-higher concentration was related to -0.045 (95% CI, -0.062, -0.027) diopter-lower cylinder. For $PM_{2.5}$, a statistically significant negative association was detected in the 3-years-before-examination exposure window with a -0.017-diopters decrease (95% CI, -0.025, -0.008) for a 10 μg/m³ concentration increase. The associations between $NO_2$, and $PM_{2.5}$ and cylinder increased with larger exposure windows, but for CO, this association reached its strongest in the 2-year exposure window and weaken thereafter. No statistically significant relationship was found between $PM_{10}$ and cylinder.

Table 1. Demographic characteristics of the study population, N = 127,709

| Characteristic | Mean (SD), n (%) | | |
| --- | --- | --- | --- |
| | 2019 | 2020 | 2021 |
| Gender | | | |
|   Male | 36194 (47.64) | 61192 (47.92) | 32037 (48.34) |
|   Female | 39786 (52.36) | 66517 (52.08) | 34237 (51.66) |
| Age (y) | 12.40 (3.13) | 13.01 (3.17) | 13.63 (3.11) |
| Type of wearing glasses | | | |
|   Non-wearing glasses | 53476 (70.38) | 92929 (72.77) | 43175 (65.15) |
| Frame glasses | 22414 (29.50) | 34749 (27.21) | 23078 (34.82) |
| Contact lenses | 26 (0.00034) | 31 (0.00024) | 21 (0.00032) |
| Orthokeratology | 64 (0.00084) | 0 | 0 |
| Education stage | | | |
| Kindergarten | 191 (0.25) | 214 (0.17) | 26 (0.039) |
| Primary school | 31361 (41.28) | 52467 (41.08) | 22509 (33.96) |
| Junior high school | 26699 (35.14) | 41521 (32.51) | 20861 (31.48) |
| Senior high school | 13559 (17.85) | 26433 (20.70) | 19505 (29.43) |
| University | 4170 (5.49) | 7074 (5.54) | 3373 (5.09) |
| School property | | | |
| Public | 60632 (79.80) | 124792 (97.72) | 65220 (98.41) |
| Private | 15348 (20.20) | 2917 (2.28) | 1054 (1.59) |
| School type | | | |
| General | 73948 (97.33) | 122854 (96.20) | 64614 (97.50) |
| Special | 0 | 6 (0.0047) | 8 (0.012) |
| Vocational | 1853 (2.44) | 4668 (3.66) | 1652 (2.49) |
| Sports | 179 (0.24) | 181 (0.14) | 0 |
| Cylinder (diopter) | -0.81 (0.88) | -0.81 (0.88) | -0.79 (0.88) |

**Table 2.** The results of panel data model with random effects for different exposure windows of each 0.1 mg/m$^3$ increase in CO, and each 10 μg/m$^3$ increase in NO$_2$, PM$_{2.5}$ and PM$_{10}$.

| Pollutant | Exposure Window | Parameter | p-value | 95% CI |
|---|---|---|---|---|
| CO | Current | -0.007 | 0.022 | (-0.013, -0.001) |
| | P1 | -0.007 | 0.045 | (-0.014, -0.002) |
| | P2 | -0.009 | 0.0011 | (-0.014, -0.004) |
| | P3 | -0.005 | 0.33 | (-0.013, -0.005) |
| $NO_2$ | Current | -0.018 | 0.00 | (-0.020, -0.015) |
| | P1 | -0.027 | 0.0001 | (-0.039, -0.014) |
| | P2 | -0.033 | 0.00 | (-0.037, -0.029) |
| | P3 | -0.045 | 0.00 | (-0.062, -0.027) |
| $PM_{2.5}$ | Current | -0.007 | 0.45 | (-0.026, 0.012) |
| | P1 | -0.009 | 0.23 | (-0.024, 0.006) |
| | P2 | -0.015 | 0.068 | (-0.030, 0.001) |
| | P3 | -0.017 | 0.0001 | (-0.025, -0.008) |
| $PM_{10}$ | Current | 0 | 0.97 | (-0.014, 0.014) |
| | P1 | 0.002 | 0.75 | (-0.011, 0.015) |
| | P2 | -0.002 | 0.89 | (-0.031, 0.027) |
| | P3 | 0 | 0.98 | (-0.020, 0.020) |

Note: Current, average concentration of the year of the exam; P1, average concentration of 1 year before the exam; P2, average concentration of 2 years before the exam; P3, average concentration of 3 years before the exam; Column Parameter refered to the values of $\beta_{0d}$ in Equation (1); Our model included school type (general, special, sports or vocational), educational stage (kindergarten, primary school, junior high school, senior high school and university), school property (public or private) and type of wearing glasses (frame

glasses, contact lenses, orthokeratology or non-wearing glasses) as covariates; CO, carbon monoxide; $NO_2$, nitrogen dioxide; $PM_{2.5}$, particulate matter less than 2.5 μm in aerodynamic diameter; $PM_{10}$, particulate matter less than 10 μm in aerodynamic diameter.

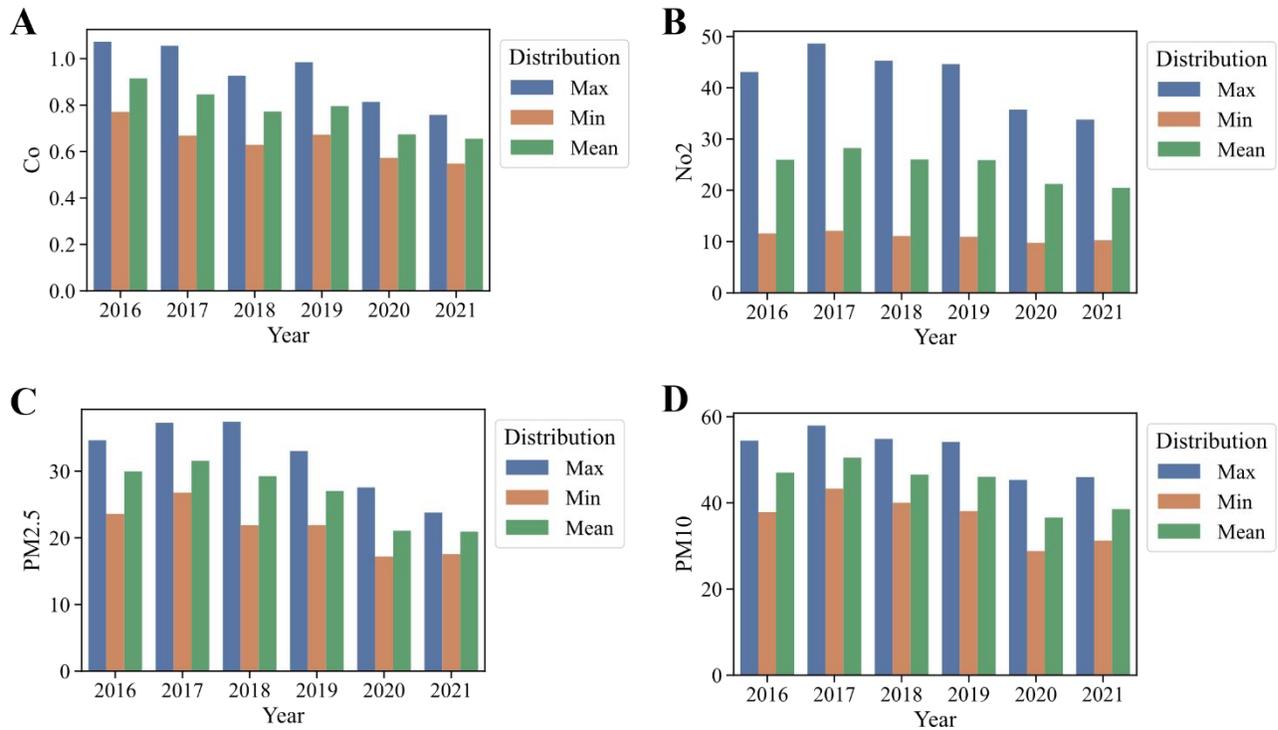

**Figure 2.** The concentration distribution of CO, $NO_2$, $PM_{2.5}$ and $PM_{10}$ containing maximum, minimum and mean values from 2016 to 2021. Note: CO, carbon monoxide; $NO_2$, nitrogen dioxide; $PM_{2.5}$, particulate matter less than 2.5 μm in aerodynamic diameter; $PM_{10}$, particulate matter less than 10 μm in aerodynamic diameter.

In our sensitivity analyses, we used specific exposure windows (P2) for CO, $NO_2$, and $PM_{2.5}$ based on their associations shown in Table 2. $PM_{10}$ was not considered due to lack of correlation. The results, displayed in Figure 3, showed significant associations between $NO_2$ and gender subgroups, as well as the age groups of 4-10 and 10-13 years old. The direction of association aligned with the baseline. However, for the age groups of 13-15 and 15-17 years old, the correlation changed and was not statistically significant. The negative impact of CO on astigmatism was observed across all subgroups, but only the male subgroup and the age group of 13-15 years old showed statistical significance. The association of $PM_{2.5}$ with astigmatism was consistent with the baseline in all subgroups except the 15-17 years old. However, significant differences were found only in the male and the age group of 13-15 years old. Overall, higher concentrations of CO, $NO_2$, and $PM_{2.5}$ were associated with lower cylinder values in children and adolescents (4-15) regardless of gender, demonstrating the robustness of our model.

**Figure 3.** Associations (95% CI) of cylinder with the average CO concentration of 2 years before examination, and the average $NO_2$ and $PM_{2.5}$ concentrations of 3 years before examination by gender subgroups and age groups. Note: CI, confidence interval; CO, carbon monoxide; $NO_2$, nitrogen dioxide; $PM_{2.5}$, particulate matter less than 2.5 μm in aerodynamic diameter; $PM_{10}$, particulate matter less than 10 μm in aerodynamic diameter; Models are adjusted for type of wearing glasses, education stage, school property, and school type; * indicates

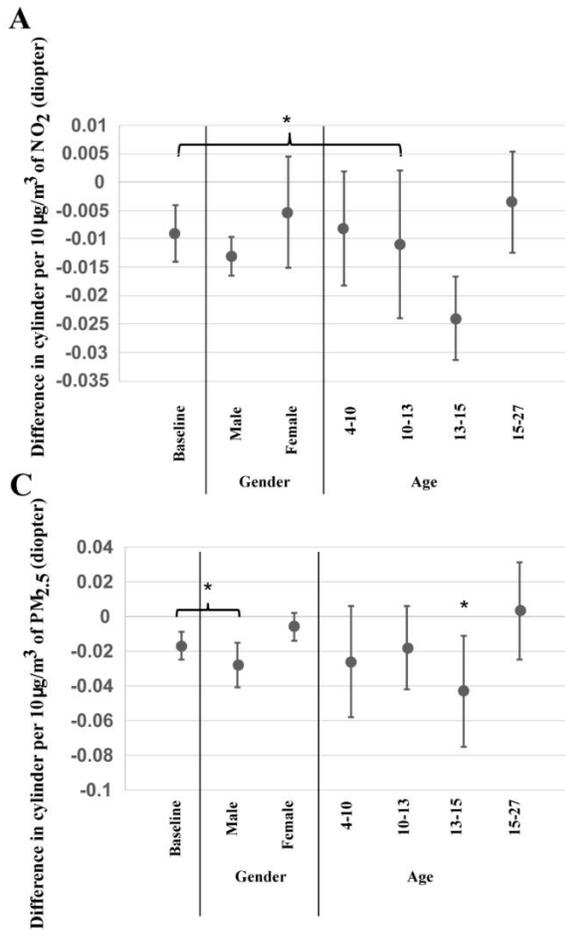
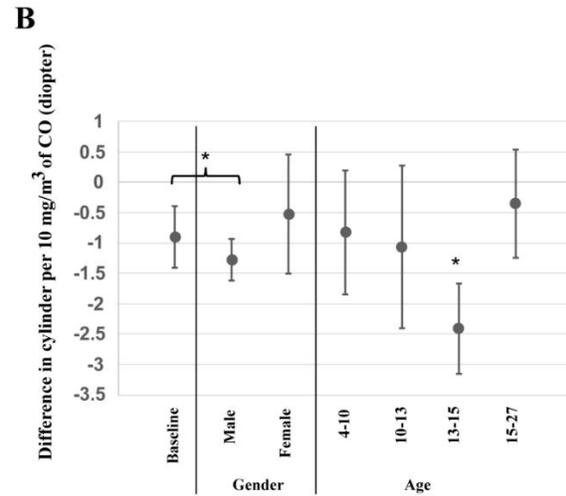

that the p-value was <.05.

## Discussion

In our large population-based longitudinal study between 2019 to 2021, we examined the influence of long-term air pollutants' exposure on the astigmatism of children and young adults from 9 cities in Guangdong Province, China, and unveiled the adverse associations between higher concentrations of CO, $NO_2$ and $PM_{2.5}$ and lower cylinder value. Our study suggests that cumulative exposure to these air pollutants would worsen astigmatism. To the best of our knowledge, this study is the first to demonstrate the adverse effects of chronic air pollution exposure on astigmatism and highlights the importance of considering longitudinal relationships.

Astigmatism is a common refractive error caused by different ocular meridians having nonuniform refractive abilities and results in multiple focus points, either in front of the retina or behind it (or both), with blurred or distorted vision. Depending on the imperfect curvature of the cornea or lens, astigmatism can be divided into corneal astigmatism and lenticular astigmatism [32]. Previous studies investigated the pathological mechanisms of air pollution on the ocular surface, such as toxicity, oxidative stress, and inflammation. Moreover, in order to adapt to the long-term changes in the environment, the ocular surface has evolved homeostatic mechanisms that make the affected people not always have symptoms [33]. A published study reported a link between $PM_{2.5}$ and tear film osmolarity, showing that the increment of air pollution raised the risk of suffering from chronic ocular surface anomalies [34]. Zhong et al. [35] found an association between greater levels of CO and $NO_2$ and a higher prevalence of dry eye disease (DED). They mentioned that an increase of 1 ppm in CO brought about an increase of 10.5-11.6% in the incidence of DED, while with each 10-ppb elevation of $NO_2$, the incidence of DED increased by 6.8–7.5%. Novaes et al. [36] pointed out that as levels of $NO_2$ rise, there was globet cell hyperplasia in the tarsal conjunctiva.

The oxidative stress and chronic inflammation caused by chronic exposure to ambient $PM_{2.5}$ and $O_3$ may damage the lens' elasticity [37, 38]. The human lens has a high degree of viscoelasticity. For the sake of seeing nearby objects more clearly, the lens can be adjusted into a thicker shape by contracting the ciliary muscles. However, the injury to the ciliary muscles and lens may cause them to lose elasticity, and it becomes difficult to contract and relax, leading to the occurrence and aggravation of myopia [39]. A study from Nepal [40] indicated that using biomass or kerosene reduced the lens's transparency, thereby elevating the possibility of being diagnosed with nuclear cataracts. They analyzed that biomass fuel smoke and tobacco smoke shared similar components, while the latter consisted of nicotine, tar, CO, etc. and was thought to be a risk factor for cataracts. Nicotine and CO can hamper the homeostasis of lipids, augment platelet aggregation, cause blood coagulation, then exacerbate circulatory disturbances, and promote the formation of atherosclerosis [41]. A population-based Korean cohort study identified a positive relationship between exposure to $NO_2$ and cataract incidence in adults aged 50 or older [42]. Given the impairments on the ocular surface and lens caused by CO, $NO_2$, and $PM_{2.5}$, it is reasonable to speculate that these air pollutants also have an effect on astigmatism.

Our study had several strengths compared to previous research. First, we analyzed a large sample size of 127,709 participants with multiple records, providing robust results. Second, we employed a longitudinal panel data model with random effects, allowing us to examine the long-term effects of air pollution, which is a novel approach. Third, we discovered that higher concentrations of CO, NO2, and PM2.5 worsened astigmatism and compared the impact across different exposure windows. However, there were limitations to our study. We could only obtain air pollution data at the city level, which limited our ability to estimate exposure at a finer resolution. We also did not consider potential bias related to variations in living environments within cities. Additionally, our study revealed associations but not causation, and there may be other factors related to air pollution that contribute to astigmatism. Further research involving cellular or animal studies could help validate the associations observed in our study.

**Acknowledgments**

We would like to thank Guangdong Eyevision Medical Technology Co., Ltd., for electronic vision screening data.